\documentclass{ifacconf}

\usepackage{graphicx} % for pdf, bitmapped graphics files
\usepackage{natbib}
\usepackage{color}                                                 % needed if you want to
\usepackage{amsmath} % assumes amsmath package installed
\usepackage{amssymb}  % assumes amsmath package installed
\usepackage{amscd}
\newtheorem{Definition}{Definition}
\newtheorem{Theorem}{Theorem}
\newtheorem{pro}{Proposition}
\newtheorem{corollary}{Corollary}

\newcommand{\sgn}{\text{sgn}}

\newcommand{\diag}{\operatorname{diag}}

\renewcommand{\S}{\mathcal{S}}

\newcommand{\R}{\mathbb{R}}

\renewcommand{\(}{\left (}
\renewcommand{\)}{\right )}
\renewcommand{\[}{\left [}
\renewcommand{\]}{\right ]}

\newcommand{\1}{{\bf 1}}

\renewcommand{\sgn}{\operatorname{sgn}}
\newcommand{\abs}{\operatorname{abs}}

\renewcommand{\qed}{\hfill\blacksquare}

\begin{document}
\begin{frontmatter}

\title{
Optimal actuator design  for minimizing the worst-case control energy}

\author[First]{Xudong Chen} 
\author[Second]{M.-A. Belabbas}

\address[First]{Department of ECEE, University of Colorado at Boulder, 
   \\Boulder, CO 80309 USA (e-mail: xudong.chen@colorado.edu).}
\address[Second]{Coordinated Science Laboratory, University of Illinois at Urbana-Champaign, 
   Urbana, IL 61801 USA (e-mail: belabbas@illinois.edu).}

\begin{abstract}
We consider the actuator design problem for linear systems. Specifically, we aim to identify an actuator which requires the least amount of control energy to drive the system from an arbitrary initial condition to the origin in the worst case. Said otherwise, we investigate the minimax problem  of minimizing the control energy over the worst possible initial conditions. Recall that  the least amount of control energy  needed to drive a linear controllable system from  any initial condition on the unit sphere to the origin is upper-bounded by the inverse of the smallest eigenvalue of the associated controllability Gramian, and moreover, the upper-bound is sharp. The minimax problem can be thus viewed as the optimization problem of minimizing the upper-bound via an actuator design. 
In spite of its simple and natural formulation, this problem is difficult to solve. In fact, properties such as the stability of the system matrix, which are not related to controllability, now play important roles.  We focus in this paper on the special case where the  system matrix is  positive definite (and hence the system is completely unstable). Under this assumption, we are able to provide a complete solution to the optimal actuator design problem and highlight the difficulty in solving the general problem. 
 %In particular, we characterize the optimal design  of the actuators for minimizing the worst-case control energy, and the corresponding  amount of energy needed for driving the system to the origin (i.e., the sharp upper-bound).  
%Even though the assumption we made for the system matrix simplifies the problem, the analysis required to obtain the optimal actuators we take in the paper is not trivial at all. The properties established along with the analysis shed light on the minimax problem in the general context.          
\end{abstract}

                       
\end{frontmatter}

\section{Introduction}
We consider in this paper the following single-input linear control system 
\begin{equation}\label{eq:LinSys}
\dot x(t) = Ax(t) + bu(t),  \hspace{20pt} x(t) \in \R^n,   
\end{equation}  
together with an infinite horizon quadratic cost function: $$\eta := \int^\infty_{0} u(t)^\top u(t) dt,$$ 
which penalizes the energy consumption for driving system~\eqref{eq:LinSys} from an initial condition $x_0$ to the origin. It is well known that if system~\eqref{eq:LinSys} is controllable (\cite{brockett1970finite}), i.e., the controllability matrix ${\rm C}(A, b) := [b,Ab,\ldots, A^{n-1} b]$ is non-singular,  then the minimal energy consumption (with respect to the initial condition $x_0$) is given by 
\begin{equation}\label{eq:DefJ}
\eta_{\min}(x_0, b) = x_0^\top W_A(b)^{-1} x_0,  
\end{equation}
where $W_A(b)$ is given by
\begin{equation}\label{eq:controllabilitygramian}
W_A(b) := \int^\infty_{0} e^{-At} bb^\top  e^{-A^\top t} dt, 
\end{equation} 
We note here that if we replace $-A$ with $A$ in~\eqref{eq:controllabilitygramian}, then the resulting matrix $W_A(b)$ is the {\em controllability Gramian} associated with~\eqref{eq:LinSys}. 
From~\eqref{eq:DefJ}, it should be clear that if an initial condition $x'_0$ (resp. the actuator $b'$) is a scalar multiple of $x_0$ (resp. $b$), i.e., $x'_0 = \alpha x_0$ (resp. $b' =\beta b$), then  $$\eta_{\min}(x'_0, b') = \frac{\alpha^2}{\beta^2}\eta_{\min}(x_0, b).$$
Following the fact, we normalize in the sequel the initial condition $x_0$, as well as the actuator $b$, such  that  $\|x_0\| =\|b\| =  1$. We further note that if the matrix $A$ is {\em stable}, i.e., all the eigenvalues of $A$ lie in the left half plane, then one can set $u(t) \equiv 0$, and hence $\eta_{\min}(x_0, b) = 0$ for any pair $(x_0,b)$. On the other hand, if the matrix $A$ is such that {\em any} of its eigenvalues lies in the right half plane (or equivalently, $-A$ is stable), then the open-loop system~\eqref{eq:LinSys} is {\em  unstable}. In particular, there does not exists a pair $(x_0,b)$  such that $\eta_{\min}(x_0,b)= 0$. For the reason mentioned above, we will assume in the sequel that system~\eqref{eq:LinSys} is open-loop unstable. We call the system matrix $A$ {\em completely unstable}.   

{\bf Problem formulation.} In this paper, we optimize $b$, with $\|b\| = 1$ fixed, for minimizing the energy consumption $\eta_{\min}(x_0,b)$ whereas the initial condition~$x_0$ is chosen so as to maximize $\eta_{\min}(x_0,b)$ for a fixed $b$. More specifically, we investigate the following {\em minimax}  problem:
\begin{equation}\label{eq:probform}
\begin{array}{l}
\phi  :=  \min_b \max_{x_0}x_0^\top W_A(b)^{-1} x_0, \vspace{3pt}\\
 \hspace{5pt}{\rm s.t.}\hspace{5pt} \|x_0\| = 1 \hspace{5pt} {\rm and} \hspace{5pt} \|b\| = 1.\\
\end{array}
\end{equation}  
For an arbitrary real symmetric matrix $M$, we denote by $\lambda_{\max} M$ (resp. $\lambda_{\min}M$) the largest (resp. smallest) eigenvalue of $M$. 
Note that the matrix  $W_A(b)$  is {\em positive semi-definite} (and hence symmetric).  
Thus, for a fix vector $b\in S^{n-1}$ where $S^{n-1}$ denotes the unit sphere in $\R^n$, we have
\begin{equation}\label{eq:transfer}
\max_{x\in S^{n-1}} x^\top W_A(b)^{-1} x = \lambda_{\max} W_A(b)^{-1} = \(\lambda_{\min}W_A(b)\)^{-1}, 
\end{equation} 
In the case $W_A(b)$ is singular, we set $\lambda_{\max} W_A(b)^{-1}$ to be infinity. From~\eqref{eq:transfer}, we obtain 
$$
 \phi = \min_{b\in S^{n-1} } \max_{x\in S^{n-1}}  x^\top W_A(b)^{-1} x = \min_{b\in S^{n-1}} \lambda_{\max} W_A(b)^{-1}.
$$
Said in another way,  the original minimax problem can be also viewed as an {\em optimization} problem that minimizes the largest eigenvalue of $W_A(b)^{-1}$ (or equivalently, maximizes the smallest eigenvalue of $W_A(b)$) over $b\in S^{n-1}$. We further define $\arg \phi$ to be the set of pairs $(x, b)$ in $S^{n-1}\times S^{n-1}$ satisfying the following properties: 
\begin{enumerate}
\item For the vector $b$, we have $\lambda_{\max}W(b)^{-1} = \phi$, i.e., the choice of $b$ minimizes $\lambda_{\max} W(b)^{-1}$.  
\item The vector $x$ is an eigenvector of $W(b)^{-1}$ corresponding to its largest eigenvalue, and hence
$$x^\top W(b)^{-1} x = \phi.$$     
\end{enumerate} 
Our objective is thus to compute both $\phi$ and $\arg\phi$.  
  
%Now, for ease of notation, we let
%$$\phi :=  \min_{b\in S^{n-1} } \max_{x\in S^{n-1}}  x^\top W(b)^{-1} x.$$ 
  
There have been a few studies in recent years related to the general problem of actuator design, and its dual sensor design for minimal actuator energy problems. For example, it has been investigated  in~\cite{Belabbas:2016kq} about how to place an actuator of system~\eqref{eq:LinSys} so as to minimize an infinite-horizon quadratic cost function: 
$$\eta = \lim_{T\to\infty}\frac{1}{T} \int_0^T (x^\top Q x+u^\top u) dt.$$ 
However, the initial condition $x_0$ there is {\em not} chosen to be in the worst case, but rather treated as a random variable drawn from a rotationally invariant distribution.  A fairly complete solution was provided there, with the assumptions that the system matrix~$A$ is stable and that the norm of the actuator $\|b\|$ is not too large. There are other investigations into this problem, most of them are  ad-hoc or application specific (see, for example~\cite{HIRAMOTO20001057,chen2014fluid,optimal_placement_geneticair1991}). Besides the works on the finite dimensional linear systems,  there has also been ample work on the infinite-dimensional case: we refer to~\cite{optimal_Actuator_morris2011} and references therein. For the problems which are specific about minimizing control energy, we mention the general investigation of various control energy measures in~\cite{bullo2014}. We also note that in the work~\cite{olshevsky2015eigenvalue}, the author there considered a similar problem for discrete-time dynamical systems, and established bounds on the smallest eigenvalue of the corresponding discrete-time controllability Gramian, and the work~\cite{dhingra2014admm} in which the authors use an L1 optimization approach to promote sparsity of a controller in  related scenarios.

{\bf Outline of contribution.}   We solve in the paper the minimax problem~\eqref{eq:probform} for the case where the system matrix $A$ is {\em positive definite}, i.e., $A$ is a symmetric matrix with positive eigenvalues. We compute explicitly the value of $\phi$. Furthermore, we provide a {\em complete} characterization of  the set $\arg\phi$, i.e., we solve the optimal actuators, as well as the corresponding worst-case initial conditions.  Even though we have  made the assumption that $A$ is symmetric so as to simplify the problem (as we will see at the  beginning of Section~2, it suffices to consider the case where $A$ is a diagonal matrix), the analysis needed for solving the minimax problem is not trivial at all. Indeed, the properties we establish for the set $\arg\phi$  provide many insights for solving the minimax problem within a general context (i.e., $A$ is arbitrary). For example, we show in the paper that if a pair $(x,b)$ lies in $\arg\phi$, then the signs of the entries of $x$ and $b$ exhibits an interlacing pattern:
$$
\sgn(x) = \pm\begin{bmatrix}
-1 & & & \\
& 1 & &\\
& & \ddots & \\ 
& & & (-1)^{n}
\end{bmatrix} \sgn(b)
$$    
where $\sgn(\cdot )$ denotes the sign function applying on a vector  entry-wise (a precise definition is given in the next section).
Such a property has also been observed via simulations for general cases where $A$ is not necessarily symmetric.  

The remainder of the paper is organized as follows: In section~2, we first introduce definitions and certain key notations, and then establish the main result, Theorem~1, of the paper, in which we completely solve $\phi$ and $\arg\phi$. Then, in section~3, we establish various properties (e.g., the interlacing sign pattern for a pair $(x,b)\in \arg\phi$) that are needed to prove the main result. We provide conclusions and outlooks. The paper ends with an appendix which contains a proof of a technical result.

\section{Solution of the minimax problem for symmetric, completely unstable systems}\label{ssec:probform}
We assume that the system matrix $A$ is positive definite, and denote by  $\lambda_1,\ldots, \lambda_n$ its eigenvalues. We further take a generic assumption that the eigenvalues of $A$ are pairwise distinct.  We re-arrange the order of the $\lambda_i$'s  so that  $$0 < \lambda_1 < \cdots < \lambda_n.$$  
Now, let $\Theta$ be the orthogonal matrix that diagonalizes~$A$, i.e., $A = \Theta \Lambda \Theta^{\top}$, with $\Lambda$ a diagonal matrix given by
$
\Lambda := \diag(\lambda_1,\ldots, \lambda_n)
$. 
Note that if we let $x' := \Theta^\top x$ and $b' := \Theta^\top b$, then by computation, 
$$
x^\top W_A(b)^{-1} x = x'^\top W_{\Lambda}(b')^{-1} x'. 
$$
This, in particular implies that 
\begin{multline*}
 \min_{b\in S^{n-1} } \max_{x\in S^{n-1}}  x^\top W_A(b)^{-1} x = \\  \min_{b'\in S^{n-1} } \max_{x'\in S^{n-1}} x'^\top W_{\Lambda}(b')^{-1} x'.   
\end{multline*}
Thus,  by following this fact,  we can assume without loss of any generality that the matrix $A$ is itself a diagonal matrix, i.e., $ A = \Lambda$. 
We will take such an assumption  for the remainder of the paper. Further,  
for ease of notation, we will suppress the sub-index $A$ of the matrix $W_A(b)$, and simply write $W(b)$.

We will now describe the main result of the paper. %  $\phi$ and $\arg\phi$. 
To proceed, we first introduce some definitions and notations. Let $\1\in \R^n$ be a vector of all ones. We define a positive definite matrix $\Psi$ as follows:
$$
\Psi := W(\1) = \int^\infty_{0} e^{-At} \1 \1^\top e^{-A^\top t} dt 
$$
Since $A = \diag(\lambda_1,\ldots, \lambda_n)$, we obtain  by computation that $\Psi$ is a Cauchy matrix (\cite{schechter1959inversion}) given by
\begin{equation}\label{eq:defPsi}
\Psi = \[ \frac{1}{\lambda_i + \lambda_j}\]_{ij},
\end{equation}
i.e., $1/(\lambda_i + \lambda_j)$ is the $ij$-th entry of $\Psi$.  
Note that with the matrix $\Psi$ defined above, one can express $W(b)$ as follows:
$$
W(b)= \diag(b) \Psi \diag(b).
$$ 
We further need the following definition:
%Now, suppose that the matrix $A$ has $k$ negative eigenvalues and $(n - k)$ positive eigenvalues. Without loss of generality, we assume that the first $k$ eigenvalues $\lambda_1,\ldots, \lambda_k$ are negative. We establish below the following result:

%\begin{pro}\label{pro:negative}
%Suppose that $(x, b)$ is a pair in $\arg\phi$; then $x$. 
%\end{pro}

%From Proposition~\ref{negative}, it suffices for us to investigate the case where all the eigenvalues of $A$ are positive, and we will assume in the sequel that $0 < \lambda_1 \cdots < \lambda_n$. To proceed, we first have some definitions and notations. Let ${\bf 1}$ be a vector of  

\begin{Definition}[Signature matrix] A real $n\times n$ matrix $M$ is a {\bf signature matrix} if it is a diagonal matrix, and the absolute value of each diagonal entry is one. 
\end{Definition}

We denote by 
$\Sigma$ the set of all $n\times n$ signature matrices, and $\sigma$ an element in $\Sigma$. It should be clear that $\Sigma$ is a finite set, with $2^n$ diagonal matrices in total. 
Among these signature matrices, there is a special signature matrix $\sigma_*$ of our particular interest, which is defined as follows: 
\begin{equation}\label{eq:defsign}
\sigma_* := 
\begin{bmatrix}
-1 & & & \\
& 1 & &\\
& & \ddots & \\ 
& & & (-1)^{n}
\end{bmatrix}.
\end{equation}
Note that the diagonal entries of $\sigma_*$ exhibit an interlacing sign pattern. 
With the definitions and notations above, we are now in a position to state the main result of the paper:

\begin{Theorem}\label{thm:mainthm}
The following hold for $\phi$ and $\arg\phi$: 
\begin{enumerate}
\item Let $\Psi$ and $\sigma_*$ be defined in~\eqref{eq:defPsi} and~\eqref{eq:defsign}, respectively. Then, 
$
\phi = \1^\top \sigma_* \Psi^{-1} \sigma_* \1
$.
\item The entries of the vector~$\sigma_* \Psi^{-1} \sigma_* \1$ are positive. Let $v_*$ be the (unique) vector of positive entries such that  
\begin{equation}\label{eq:definevstar}
\begin{bmatrix}
v^2_{*,1} \\
\vdots \\
v^2_{*,n}
\end{bmatrix} = \frac{ 1}{\phi}\sigma_* \Psi^{-1} \sigma_* \1.
\end{equation}
Then, $\arg\phi$ has $2^{n + 1}$ elements:
\begin{equation}\label{eq:argvarphi}
\arg\phi = \{ (\pm \sigma  v_*, \sigma\sigma_{*} v_*) \mid \sigma\in \Sigma\}.
\end{equation}
\end{enumerate}\,
\end{Theorem}

The remainder of the section is devoted to the proof of Theorem~\ref{thm:mainthm}. We will first establish in Subsection~\ref{ssec:signpattern} the sign pattern for a pair $(x, b)\in \arg\phi$. Then, in Subsection~\ref{ssec:potentialfunction}, we reformulate the minimax problem as an optimization problem maximizing $\lambda_{\min}W(b)$, and establish a necessary and sufficient condition for a vector $b\in S^{n-1}$ to be a critical point of the  function $\lambda_{\min}W(b)$. A complete proof of Theorem~\ref{thm:mainthm} is provided in Subsection~2.3. 
%In Subsections~\ref{ssec:signpattern} and~\ref{ssec:proofofthm1}, we establish various properties that are necessary to prove Theorem~\ref{thm:mainthm}. Note that from~\eqref{}
%In Subsection~\ref{ssec:signpattern} 

\subsection{Sign patterns of optimal solutions}\label{ssec:signpattern}
 %To proceed, we first define a function $\varphi (x, b)$, for $x\in \R^n$ and $b\in\R^{n}$ as follows: 
 %$$
 %\varphi(x, b) = x^\top W(b) x. 
 %$$
% We then let 
% $$
%\varphi_*  :=  \max_{B\in S^{n-1}} \min_{x\in S^{n-1}} \varphi(x, b). 
% $$
%Recall that we have defined $\arg \phi$ to be the set of pairs $(x, b)$ in $S^{n-1} \times S^{n-1}$ such that $\phi(x,b) = \phi$. Similarly, we define  
% $$
 %\arg \varphi_*:= \{ (x, b)\in S^{n-1}\times S^{n-1} \mid \varphi(x, b) = \varphi_*\}.
 %$$
% With the definition above, we have the following fact:

%\begin{lem}\label{lem:inverse}
%The two sets $\arg \phi$ and $\arg \varphi_*$ coincide with each other, and moreover,  $\phi = \varphi^{-1}_* $. 
%\end{lem}

%\begin{pf}
%$\qed$\end{pf}
 
%From Lemma~\ref{lem:inverse}, it suffices to compute $\arg \varphi_*$. In particular, we will prove toward the end the section that $Z = \arg \varphi_*$. 

To proceed, we first have some definitions and notations. We denote by $\sgn(\cdot)$ the sign function, i.e.,  for a scalar $x\in \R$, we let 
$$\sgn(x):=
\left\{
\begin{array}{ll}
1  & x> 0, \\
0 & x = 0, \\
-1 & x< 0.
\end{array}
\right.$$ 
Then, for an arbitrary matrix $A = [a_{ij}]_{ij}\in \R^{m\times n}$ (or simply a vector), we define $\sgn(A)$ by letting $\sgn$ act on each entry of $A$:  $$\sgn(A):= \[\sgn(a_{ij})\]_{ij}.$$ 
We further let ${\rm abs}(A)$ be a positive matrix defined by replacing each entry of $A$ with its absolute value. We note that if $v$ is a vector, then $\abs(v)$ can be expressed as follows:
$$
\abs(v) = \diag(\sgn(v)) v.
$$ 
With the definitions above,  we establish below the following result:  

\begin{pro}\label{pro:signpattern} 
Let $(x_*, b_*)$ be a pair in $\arg\phi$. Then, all the entries of $x_*$ and of $b_*$ are nonzero. Moreover, 
\begin{equation}\label{eq:signrelation}
\sgn(x_*) = \pm \sigma_* \sgn(b_*),
\end{equation}
where $\sigma_*$ is the signature matrix defined  in~\eqref{eq:defsign} 
\end{pro}

We establish below proposition~\ref{pro:signpattern}. 
First, recall that the matrix $\Psi\in \R^{n\times n}$, defined in~\eqref{eq:defPsi}, is positive definite. It then follows that $$W(b) = \diag(b)\Psi\diag(b)$$ is positive semi-definite, and is positive definite if and only if the entries of $b$ are all nonzero. Indeed, the number of zero eigenvalues of $W(b)$ is the same as the number of zero entries of $b$.  
We also note that for a fixed vector $b\in \R^n$, 
$$
\max_{x\in S^{n-1}} x^\top W(b)^{-1}x = \lambda_{\max}W(b)^{-1}.$$  Hence, if $b_1, b_2\in S^{n-1}$ are two vectors such that $b_1$ has no zero entry and $b_2$ has at least one zero entry, then 
$$
\lambda_{\max} W(b_1)^{-1}  < \lambda_{\max} W(b_2)^{-1}  = \infty. 
$$
The arguments above then imply the following fact:

\begin{lem}\label{lem:entryofb}
If $(x_*,b_*)$ is in $\arg\phi$, then all the entries of $b_*$ are nonzero.  
\end{lem}

Now, following Lemma~\ref{lem:entryofb},  we fix  a vector $b\in \R^n$ with no zero entry. It then follows that the matrix $W(b)$ is invertible, with its inverse given by
$$
W(b)^{-1} = \diag(b)^{-1} \Psi^{-1} \diag(b)^{-1}. 
$$ 
%Note that if $x\in \R^n$ is an eigenvector of $W(b)$, with $\lambda > 0$ the smallest eigenvalue of $W(b)$,  then $x$ is an eigenvector of $W(b)$, with $\lambda^{-1}$ the largest eigenvalue of $W(b)^{-1}$. 
We now compute explicitly $\Psi^{-1}$. To do so, we first recall a relevant fact about the determinant of a Cauchy matrix:

\begin{lem}\label{lem:usefulfact}
Let $\{\alpha_i\}^n_{i = 1}$ and $\{\beta_i\}^n_{i = 1}$ be two sets of positive numbers, and  $M$ be an $n\times n$ Cauchy matrix:
$$
M := \[\frac{1}{\alpha_i + \beta_j}\]_{ij}.
$$ 
Then, the determinant of $M$ is given by
\begin{equation}\label{eq:computedetm}
\det(M) = \prod^n_{k = 1}\frac{1}{\alpha_k + \beta_k} \prod_{1\le i< j\le n} \(\frac{\alpha_j - \alpha_i}{\alpha_i + \alpha_j}  \frac{\beta_j - \beta_i}{\beta_i + \beta_j}\). 
\end{equation}
In particular, $\det(M) > 0$.
\end{lem}

This fact has certainly been observed before (see, for example,~\cite{schechter1959inversion}). We reproduce a proof in the Appendix for the sake of completeness. 
 
With the lemma at hand, we return to  the computation of $\Psi^{-1}$: First, we write
$$
\Psi^{-1} = \frac{1}{\det(\Psi)} {\rm Co}(\Psi)^\top,
$$ 
where ${\rm Co}(\Psi) = {\rm Co}(\Psi)_{ij}$ is the cofactor of $\Psi$:
$$
{\rm Co}(\Psi)_{ij} = (-1)^{i+j} \det\(M_{ij}\),
$$ with $M_{ij}$ a minor of $\Psi$ obtained by  deleting the $i$-th row and the $j$-th column of $\Psi$. 
Since $\Psi$ is a Cauchy matrix, we use~\eqref{eq:computedetm} to obtain
$$
\det(\Psi) = \prod^n_{k =1}\frac{1}{2\lambda_k} \prod_{1\le i< j \le k} \(\frac{\lambda_{j} -\lambda_i}{\lambda_i + \lambda_j}\)^2
$$
We further note that each minor $M_{ij}$ is also a Cauchy  matrix. Thus, using~\eqref{eq:computedetm} again, we have 
\begin{multline*}
\det\(M_{ij}\) = (-1)^{i+j}\frac{\prod^{n}_{k = 1}2\lambda_k}{\prod^j_{k = i} 2\lambda_k} \prod^{j-1}_{k = i}(\lambda_{k+1} + \lambda_k) \\ 
 \frac{\prod_{1\le i'< j' \le n} \(\frac{\lambda_{j'} -\lambda_{i'}}{\lambda_{i'} + \lambda_{j'}}\)^2}{\prod_{k \neq i} \frac{\lambda_{i} - \lambda_{k}}{\lambda_i + \lambda_k}
\prod_{k\neq j} \frac{\lambda_{j} - \lambda_{k}}{\lambda_j + \lambda_k}  
}.
\end{multline*}
In particular, $\det(M_{ij}) > 0$. 
Combining the computations above, we have the following fact for $\Psi^{-1}$: %as a corollary to Lemma~\ref{lem:usefulfact}:  

\begin{lem}\label{cor:inverseofsigma}
Let $\Psi^{ij}$ be the $ij$-th entry of $\Psi^{-1}$, with $i\le j$. Then,  
\begin{equation}\label{eq:computeinverse}
\Psi^{ij} = \frac{ \prod^j_{k = i} 2\lambda_{k}}{\prod^{j-1}_{k = i}(\lambda_{k+1} + \lambda_k)}\prod_{k \neq i} \frac{\lambda_{i} + \lambda_{k}}{\lambda_i - \lambda_k}
\prod_{k\neq j} \frac{\lambda_{j} + \lambda_{k}}{\lambda_j - \lambda_k}.  
\end{equation}
In particular, $\Psi^{-1}$ has the checkerboard sign pattern: 
$$
\sgn\(\Psi^{-1} \) = \[ (-1)^{i+j}\]_{ij}.
$$\,
%Note that if $i = j$, then by default, we set 
%$$
%\prod^{j-1}_{k= 1} \lambda_{k+1} + \lambda_{k} = 1. 
%$$
\end{lem}

We recall that a square matrix $M$ is said to be {\bf irreducible} if there does not exist a permutation matrix $P$ such that $P M P^\top$ is a block upper triangular matrix. We also recall from the Perron-Frobenius theorem (see, for example,~\cite{gantmakher1998theory}) that if $M$ is a irreducible matrix of positive entries, then $M$ has a unique largest eigenvalue~$\lambda$.   Furthermore, if $v$ is an eigenvector of $M$ corresponding to the eigenvalue~$\lambda$, then  by appropriate scaling, one has that $\|v\| = 1$ and $\sgn(v) = \1$. 
The following fact is then an immediate consequence of Lemma~\ref{cor:inverseofsigma}:

\begin{corollary}
The matrix $\Psi$ is positive definite. It has a unique largest eigenvalue $\lambda_{\max}$ and a unique smallest eigenvalue $\lambda_{\min}$. Moreover, if we let $v_{\max}$ (resp, $v_{\min}$) be the eigenvector of $\Psi$ corresponding the eigenvalue $\lambda_{\max}$ (resp. $\lambda_{\min}$), then by appropriate scaling, we have
$$
\sgn(v_{\max}) = \1 \hspace{5pt} \mbox{ and } \hspace{5pt} \sgn(v_{\min}) = \sigma_* \1. 
$$\,
\end{corollary}

\begin{pf}
First, note that $\Psi$ is an irreducible matrix of positive entries, and hence from the Perron Frobenius theorem, it has a unique largest eigenvalue $\lambda_{\max}$, and hence we can choose $v_{\max}$ such that $\sgn(v_{\max}) = \1$. On the other hand, from Lemma~\ref{cor:inverseofsigma}, we have that  
$\sigma_* \Psi^{-1} \sigma_*$ is also an irreducible matrix of positive entries. We let $\lambda'_{\max}$ be the unique largest eigenvalue of $\sigma_* \Psi^{-1} \sigma_*$, and $u_{\max}$ be an eigenvector of $\sigma_* \Psi^{-1} \sigma_*$ corresponding to the eigenvalue $\lambda'_{\max}$ with $\sgn(u_{\max}) = \1$.  
Since $\Psi^{-1}$ is similar to $\sigma_* \Psi^{-1} \sigma_*$ via the signature matrix $\sigma_*$, we have that $\lambda'_{\max}$ is the unique largest eigenvalue of $\Psi^{-1}$, with $\sigma_* u_{\max}$ a corresponding eigenvector. Note, in particular, that $\sgn(\sigma_* u_{\max}) = \sigma_* \1$. Now, let $v_{\min}:= \sigma_* u_{\max}$. It then follows that 
$$
\Psi v_{\min}= \frac{1}{\lambda'_{\max}} v_{\min}, 
$$
with $1/\lambda'_{\max}$ the unique smallest eigenvalue of $\Psi$.
$\qed$\end{pf}

With the preliminaries results established above, we are now in a position to prove~Proposition~\ref{pro:signpattern}. 

\begin{pf*}{Proof of Proposition~\ref{pro:signpattern}.}
Let $(x_*, b_*)$ be a pair in $\arg \phi$. From Lemma~\ref{lem:entryofb}, the entries of $b_*$ are nonzero. We recall that ${\rm abs}(b_*)\in \R^n$ is defined by replacing each entry of $b_*$ with its absolute value, and 
$$
\abs(b_*) = \diag(\sgn(b_*)) b_*.
$$ 
Now, we define a matrix $M$ as follows: 
\begin{equation*}
M:= \sigma_*\diag(\abs(b_*))^{-1} \Psi^{-1} \\ \diag(\abs(b_*))^{-1}\sigma_*.
\end{equation*}  
Then, from Lemma~\ref{cor:inverseofsigma}, $M$ is an irreducible matrix of positive entries. 
%We also note that since $\Psi^{-1}$ is positive definite, so is $M$.  
Appealing to the Perron-Frobenius theorem, we know that there is a unique largest eigenvalue $\lambda$ of $M$. We further let $v$ be the eigenvector of $M$ corresponding to the eigenvalue~$\lambda$, with $\|v\| = 1$ and $\sgn(v) = \1$.

Now, recall that $W(b_*)^{-1}$ is given by:  $$W(b_*)^{-1} = \diag(b_*)^{-1} \Psi^{-1} \diag(b_*)^{-1}.$$ 
We thus obtain   
$$
M = \sigma_*\diag(\sgn(b_*))W(b_*)^{-1} \\ \diag(\sgn(b_*))\sigma_*, 
$$
In other words, $M$ and $W(b_*)^{-1}$ are related by a similarity transformation, via the signature matrix $\sigma_*\diag(\sgn(b_*))$. This, in particular, implies that the matrix $W(b_*)^{-1}$ has $\lambda$ as its {\em unique} largest eigenvalue. Moreover, if we let 
$$
u := \sigma_*\diag(\sgn(b_*))v,
$$    
then $u$ is an eigenvector of $W(b_*)^{-1}$ corresponding to the eigenvalue~$\lambda$. We further note that $$\|u\| = \|\sigma_*\diag(\sgn(b_*))v\| = 1,$$ and hence for fixed $b_*$, the only possible solutions for $x_*\in S^{n-1}$ (such that $(x_*, b_*)\in \arg \phi$) are given by
$
x_* = \pm u
$.  
Then, using the fact that $\sgn(v) = {\bf 1}$,  we conclude that  
$$
\sgn(x_*) = \pm \sgn(u) = \pm\sigma_*\sgn(b_*), 
$$
which completes the proof. 
$\qed$\end{pf*}

\subsection{A potential function and its critical points}\label{ssec:potentialfunction}   
For a vector $b\in S^{n-1}$, we let $\xi(b)$ be the smallest eigenvalue of the matrix $W(b)$. 
We think of $$\xi: b\mapsto \lambda_{\min}W(b)$$ as a potential function defined over $S^{n-1}$. 
Let $Z$ be a proper subset of $S^{n-1}$ defined by collecting any vector $v\in S^{n-1}$ with nonzero entries:  
$$
Z:= \left\{ v \in S^{n-1} \mid |v_i| > 0, \hspace{5pt}\forall\, i = 1,\ldots, n \right \}.
$$
%Next, we define a function $\xi$ over $Z$ as follows: For a vector $b\in Z$, we let $\xi(b)$ be the smallest eigenvalue of the matrix $W(b)$:
We note that for any $b\in Z$, the matrix $W(b)$ is nonsingular, and hence $\xi(b) > 0$. 
By the same arguments as used to establish Lemma~\ref{lem:entryofb}, we know that  
$
\max_{b\in S^{n-1}} \xi(b)  
$ 
can be achieved by a vector in~$Z$; indeed, the set $S^{n-1} - Z$ is comprised of the global minima of~$\xi$.  
We also note that if $b\in Z$, then from Proposition~\ref{pro:signpattern}, there exists a unique smallest eigenvalue of $W(b)$. Thus, the corresponding eigenspace is of dimension one.  
%We further note that if a vector $b_*\in Z$ maximizes~$\xi$, then $b_*$ minimizes the largest eigenvalue of $W(b_*)^{-1}$. 
Now, let $b_*\in Z$  be a global maximum point of $\xi$, and $x_*\in S^{n-1}$ be an eigenvector of $W(b_*)$ corresponding to the smallest eigenvalue of $W(b_*)$.  Then, it should be clear that  
$(x_*,b_*)\in \arg \phi$. Conversely, if $(x_*,b_*)$ is in $\arg\phi$, then $b_*$ maximizes $\xi(b)$. It thus suffices  to locate the global maxima of~$\xi$.

For a vector $b\in S^{n-1}$,  we denote by $D_b\xi$ the derivative of $\xi$ at~$b$.  The map $D_b\xi$ sends a vector $v$ in $T_b S^{n-1}$---the tangent space of $S^{n-1}$ at $b$---to a real number. The vector~$b$ is said to be a {\bf critical point} of~$\xi$ if $D_b \xi$ is identically zero, i.e., $D_b\xi(v) = 0$ for all $v\in T_b S^{n-1}$. Note that a local maximum point of $\xi$ is necessarily a critical point of~$\xi$.  
The following fact then presents a necessary condition for a vector $b\in Z$ to be a critical point of~$\xi$:
%We prove the theorem by establishing the following result:

\begin{pro}\label{pro:relationship}
Let $b\in Z$ be a critical point of $\xi$, with $\lambda := \xi(b)$ the (unique) smallest eigenvalue of $W(b)$.  Let $x\in S^{n-1}$ be an eigenvector of $W(b)$ corresponding to the eigenvalue $\lambda$. Then, the following holds:  
\begin{equation}\label{eq:relationship}
\left\{
\begin{array}{l}
W(b) x = \lambda x \vspace{3pt} \\
W(x) b = \lambda b. 
\end{array}
\right. 
\end{equation}\,
\end{pro}

We establish below Proposition~\ref{pro:relationship}. To proceed, we evaluate $D_b \xi$ for a vector $b\in Z$. 
%First, note that the first expression of~\eqref{eq:relationship}:
%$$
%W(b) x = \lambda x
%$$
%directly follows from the fact that for a fixed vector $b_*\in S^{n-1}$, $x_*$ has to be an eigenvector of $W(b_*)^{-1}$ coresponding to its largest eigenvalue, and hence an eigenvector of $W(b_*)$ corresponding to the smallest eigenvalue $\lambda$ of $W(b_*)$.  We establish below the other expression:
%$$
%W(x_*) b_* = \lambda b_*. 
%$$    
%First, note that the  function $\xi$ is analytic. 
%Recall that in the proof of Proposition~\ref{pro:signpattern},  we have shown that  for any $b \in Z$, the matrix $W(b)^{-1}$ (resp. $W(b)$) has a unique largest (resp. least) eigenvalue.  It then follows that the map $\xi$ is analytic over the set $Z$. 
First, note that the tangent space of $S_n$ at $b$ is given by  
\begin{equation}\label{eq:tangent space}
T_b S^{n-1} = \left\{ v\in \R^n \mid v^\top b = 0 \right\}.
\end{equation}   
We then note the following fact: Let $M$ be an arbitrary symmetric matrix, with $\lambda$ a distinct eigenvalue and $v\in S^{n-1}$ a corresponding eigenvector. Suppose that we perturb $M$ to $(M + \epsilon N)$ for $N$ symmetric and $\epsilon$ sufficiently small, then up to the first order of $\epsilon$, the perturbed eigenvalue $\lambda(\epsilon)$ of $(M + \epsilon N)$ is given by
$
\lambda(\epsilon) = \lambda + \epsilon v^\top N v
$.  
We further recall, from the proof of Proposition~\ref{pro:signpattern}, that for any vector $b \in Z$, the matrix $W(b)^{-1}$ (resp. $W(b)$) has a unique largest (resp. least) eigenvalue. The arguments above then imply the following fact:

\begin{lem}
Let $b\in Z$, and $v\in T_bS^{n-1}$. Then, 
\begin{equation}\label{eq:derivativeofxi}
D_b \xi (v)  = 2 v^\top W(x) b.   
\end{equation}
where $x\in S^{n-1}$ is an eigenvector of $W(b)$ corresponding to the smallest eigenvalue of $W(b)$. 
\end{lem}

\begin{pf}
The lemma follows directly from computation; indeed, if we perturb $b$ to $b + \epsilon v$, then up to the first order of $\epsilon$, the perturbed  matrix $W(b + \epsilon v)$ is given by
\begin{multline*}
W(b + \epsilon v) = W(b) + \\ \epsilon \(\diag(v) \Psi \diag(b) + \diag(b) \Psi \diag(v)\).
\end{multline*}
Since $W(b)$ has a unique smallest eigenvalue, for $\epsilon$ small,  the perturbed matrix $W(b + \epsilon v)$ also has a unique smallest eigenvalue, which is given by 
\begin{multline*}
\lambda_{\min}W(b + \epsilon v) = 
\lambda_{\min}W(b) + \\ 2 \epsilon x^\top \diag(v) \Psi \diag(b) x + {\rm o}(\epsilon)
\end{multline*}
We further note that
$$
x^\top \diag(v) \Psi \diag(b) x = v^\top W(x) b,
$$
and hence
$
D_b\xi(v) = 2v^\top W(x) b
%\begin{array}{lll}
%D_b\xi(v) & = & \displaystyle \lim_{\epsilon\to 0} \frac{\lambda_{\min}W(b + \epsilon v) - \lambda_{\min}W(b) }{\epsilon} \vspace{3pt}\\
%&  = & 2v^\top W(x) b.
%\end{array}
$.
$\qed$\end{pf}

With the preliminaries above, we are now in a position to prove Proposition~\ref{pro:relationship}:

\begin{pf*}{Proof of Proposition~\ref{pro:relationship}.}
%First, note that if $(x_*, b_*)\in \arg\phi$, then $b_*$ must be a critical point of the map $\xi$ and $x_*$ is an eigenvector of $W(b_*)$ corresponding to its smallest eigenvalue. 
Since~$b\in Z$ is a critical point of $\xi$, we have that for any vector $v\in T_{b}Z$, 
$
D_{b} \xi(v) = 0
$. From~\eqref{eq:derivativeofxi}, we have
$
v^\top W(x) b= 0
$. 
Since the expression above holds for all $v\in T_{b}S^{n-1}$, we know from~\eqref{eq:tangent space} that
$
W(x) b =\mu b
$,  
for some constant~$\mu$. 
It now suffices to show that $\mu = \lambda$. To see this, we first note that 
\begin{equation}\label{eq:bxbx1}
b^\top W(x) b = \mu b^\top b = \mu.
\end{equation}
On the other hand, we have
$
W(b) x = \lambda x 
$, 
and hence
\begin{equation}\label{eq:bxbx2}
x^\top W(b) x = \lambda x^\top x = \lambda.
\end{equation}
Since the left hand side of~\eqref{eq:bxbx1} coincides with the left hand side of~\eqref{eq:bxbx2}, we conclude that $\mu = \lambda$. %This completes the proof. 
\end{pf*} 

\subsection
{Analysis and proof of Theorem~\ref{thm:mainthm}} 
We prove here Theorem~\ref{thm:mainthm}. The proof relies on the use of  Proposition~\ref{pro:relationship}. More specifically, we prove Theorem~\ref{thm:mainthm} by establishing  the following fact as a corollary to Proposition~\ref{pro:relationship}:

\begin{corollary}\label{cor:criticalpoints}
There are $2^n$ isolated critical points of the potential function~$\xi$ over the set~$Z$. They are given by 
$
\{\sigma v_* \mid \sigma \in \Sigma\}
$, 
where $v_*$ is a positive vector defined in~\eqref{eq:definevstar}. 
Furthermore, the following properties hold: 
\begin{enumerate}
\item The function~$\xi$ holds the same value at each of these critical points: 
$$
\xi(\sigma v_*) = \frac{1}{\1 \sigma_* \Psi \sigma_* \1}.
$$
Thus, the $2^n$ critical points form the global maxima of the function~$\xi$. 
\item For a critical point $\sigma v_*$ of~$\xi$, the two vectors $\pm \sigma \sigma_* v_*$ are the eigenvectors of the matrix $W(\sigma v_*)$ corresponding to its (unique) smallest eigenvalue.  
\end{enumerate} \,
\end{corollary} 

%We  establish in this subsection Theorem~\ref{thm:mainthm}. Note that Theorem~\ref{thm:mainthm} is then a direct consequence of Propositions~\ref{pro:signpattern} and~\ref{pro:relationship}: 

\begin{pf}
%First, note that from Lemma~\ref{lem:inverse}, we can obtain $\phi$ and $\arg\phi$ by  evaluating $\varphi_*$ and $\arg\varphi_*$.   
Let $b\in Z$ be a critical point of $\xi$, $\lambda$ be the smallest eigenvalue of $W(b)$, and $x\in S^{n-1}$ be an eigenvector of $W(b)$ corresponding to the eigenvalue~$\lambda$.  Then,  from~\eqref{eq:relationship}, we have that 
\begin{equation}\label{eq:xbxb}
\left\{
\begin{array}{l}
\Psi\diag(b) x = \lambda \diag(b)^{-1}  x \vspace{3pt} \\
 \Psi\diag(x) b = \lambda \diag(x)^{-1} b, 
\end{array}
\right. 
\end{equation}
Since $\lambda \neq 0$ and 
$$\diag(b) x =  \diag(x) b,$$ 
we obtain from~\eqref{eq:xbxb} that
$$
\diag(b)^{-1}  x =  \diag(x)^{-1}  b. 
$$ 
In other words, if we let $b_{i}$ (resp. $x_{i}$) be the $i$-th entry of $b$ (resp. $x$), then the expression above implies that
\begin{equation}\label{eq:gashenguang}
|x_{i}| = |b_{i}|, \hspace{10pt} \forall i = 1,\ldots, n. 
\end{equation}
On the other hand, from Proposition~\ref{pro:signpattern}, we have 
$$
\sgn(x) = \pm \sigma_* \sgn(b). 
$$
We then combine this fact with~\eqref{eq:gashenguang}, and  obtain that
\begin{equation}\label{eq:sign1}
x = \pm \sigma_* b.
\end{equation}  
From~\eqref{eq:xbxb} and~\eqref{eq:sign1}, we then have
$$
\Psi\sigma_*
\begin{bmatrix}
b^2_{1} \\
\vdots \\
b^2_{n}
\end{bmatrix} = \lambda \sigma_* {\bf 1}, 
$$
and hence
\begin{equation}\label{eq:finalcondition}
\begin{bmatrix}
b^2_{1} \\
\vdots \\
b^2_{n}
\end{bmatrix} = \lambda \sigma_* \Psi^{-1} \sigma_* {\bf 1}. 
\end{equation}
We note that $\Psi^{-1}$ has the checkerboard sign pattern, and hence the entries of the matrix $\sigma_* \Psi^{-1} \sigma_*$ on the right hand side of~\eqref{eq:finalcondition} are positive.  

With~\eqref{eq:finalcondition} at hand, we can compute  explicitly the scalar $\lambda$ and the vector $(b^2_{1}, \ldots,  b^2_{n})$: First, for the scalar~$\lambda$,  we use the fact that $\sum^{n}_{i = 1}b^2_{i} = 1$, and obtain 
$$
\lambda =  \( {\bf 1}^\top   \sigma_* \Psi^{-1} \sigma_* {\bf 1}\)^{-1}.
$$  
It then follows that 
\begin{equation*}
\begin{bmatrix}
b^2_{1} \\
\vdots \\
b^2_{n}
\end{bmatrix}  = \frac{\sigma_* \Psi^{-1} \sigma_* {\bf 1}}{{\bf 1}^\top \sigma_* \Psi^{-1} \sigma_* {\bf 1}} = 
\begin{bmatrix}
v^2_{*,1} \\
\vdots \\
v^2_{*,n}
\end{bmatrix}. 
\end{equation*}
Note that the expression above uniquely determines the set of critical points of~$\xi$ over $Z$: There are $2^n$ critical points of $\xi$, one to one corresponding to the signature matrices:    
$$
\left \{ \sigma v_* \mid \sigma\in \Sigma\right\}.
$$
Moreover, the function~$\xi$ holds the same value at each of these critical points: 
$$
\xi(\sigma v_*) = 
\lambda = \({\bf 1}^\top \sigma_* \Psi^{-1} \sigma_* {\bf 1}\)^{-1}.
$$ 
This  establishes the first item of the corollary. The second item directly follows from~\eqref{eq:sign1}. 
$\qed$\end{pf}

\section{Summary}
We considered in the paper the actuator design problem for a linear control system so as to minimize the worst case control energy, where worst case is to be understood with respect to the initial state of the system. This problem is in general difficult, and we focussed here only on systems for which the infinitesimal generator of the dynamics $A$ is positive definite. Under this assumption, we provided a complete characterization of the optimal actuators and the corresponding worst case initial states.  We also evaluated the value of the worst-case energy needed for the optimal actuator. Along the way, we highlighted several structural properties of the set of optimal actuators and their corresponding worst case initial states, such as the interlacing sign pattern of their entries. Future work may focus on a general case where $A$ is not necessarily symmetric, and the cases where one has multiple actuators.    
%We will generalize the results developed in the paper to the case of arbitrary generator of the dynamics $A$.

\section*{Appendix}
%\setcounter{subsection}{0}
%\subsection{Proof of Lemma~\ref{lem:usefulfact}}
We prove here  Lemma~\ref{lem:usefulfact}. The proof is carried out by induction on the dimension of the matrix~$M$. For the base case, we have that $M$ is a scalar given by $1/(\alpha_1 + \beta_1)$. Thus, 
$
\det(M) = 1/(\alpha_1 + \beta_1)
$, 
and hence~\eqref{eq:computedetm} holds. 

For the inductive step, we assume that Lemma~\ref{lem:usefulfact} holds for~$n =k$, and prove for $n = k+1$. To proceed, we first partition the matrix $M$ into $2\times 2$ blocks as follows:
$$
M = 
\begin{bmatrix}
M_{11} & u\\
v^\top &  \frac{1}{\alpha_{k+1} + \beta_{k+1}}
\end{bmatrix},
$$
with $M_{11}$ a $k\times k$ matrix. Next, we have the following elementary row operations on the matrix $M$: 
$$
 \begin{bmatrix}
I_{k\times k}  & -(\alpha_{k+1} + \beta_{k+1})u \\
0 & 1
\end{bmatrix}
\begin{bmatrix}
M_{11} & u\\
v^\top &  \frac{1}{\alpha_{k+1} + \beta_{k+1}}
\end{bmatrix},
$$
by which we obtain the following matrix:
$$
\begin{bmatrix}
M_{11} - (\alpha_{k+1} + \beta_{k+1}) uv^\top  & 0 \\
v^\top & \frac{1}{\alpha_{k+1} + \beta_{k+1}}
\end{bmatrix}
$$ 
Note that the elementary row operation defined above does not change the determinant of $M$, and hence  
\begin{equation}\label{eq:detM}
\det(M) = \frac{\det\(M_{11} - (\alpha_{k+1} + \beta_{k+1}) uv^\top\)}{\alpha_{k+1}+ \beta_{k+1}}. 
\end{equation}   
It thus suffices to evaluate the determinant of the matrix $M_{11} - ({\alpha_{k+1}+ \beta_{k+1}}) uv^\top$.  

To do so, we first obtain, by computation, the following expression:
\begin{equation}\label{eq:threematrices}
M_{11} - ({\alpha_{k+1}+ \beta_{k+1}}) uv^\top = D_\alpha M' D_{\beta},
\end{equation}  
where $M'\in \R^{k\times k}$ is given by
$$
M' := \[\frac{1}{\alpha_i + \beta_j}\]_{1\le i, j\le k},
$$
and $D_\alpha, D_{\beta} \in \R^{k\times k}$ are diagonal matrices given by
$$
D_{\alpha} := 
\begin{bmatrix}
\frac{\alpha_{k+1} - \alpha_1}{\alpha_{k+1} + \alpha_1}  & & \\
& \ddots & \\
& & \frac{\alpha_{k+1} - \alpha_k}{\alpha_{k+1} + \alpha_k}
\end{bmatrix}
$$
and
$$ 
D_{\beta} := 
\begin{bmatrix}
\frac{\beta_{k+1} - \beta_1}{\beta_{k+1} + \beta_1}  & & \\
& \ddots & \\
& & \frac{\beta_{k+1} - \beta_k}{\beta_{k+1} + \beta_k}
\end{bmatrix}.
$$ 
From~\eqref{eq:detM} and~\eqref{eq:threematrices}, we have 
\begin{equation}\label{eq:eq1}
\det(M) = \frac{\det(M')}{\alpha_{k+1} + \beta_{k+1}} \prod^k_{i = 1}\(\frac{\alpha_{k+1} - \alpha_i}{\alpha_{k+1} + \alpha_i} 
\frac{\beta_{k+1} - \beta_i}{\beta_{k+1} + \beta_i}\).
\end{equation}
We then appeal to the induction hypothesis and  obtain the determinant of $M'$: 
\begin{equation}\label{eq:eq2}
\det(M') = \prod^k_{i = 1}\frac{1}{\alpha_i + \beta_i} \prod_{1\le i< j\le k} \(\frac{\alpha_j - \alpha_i}{\alpha_i + \alpha_j}  \frac{\beta_j - \beta_i}{\beta_i + \beta_j}\).
\end{equation}
Combining~\eqref{eq:eq1} and~\eqref{eq:eq2}, we conclude that~\eqref{eq:computedetm} holds. This completes the proof. 
\hfill{$\qed$}

%\subsection{Proof of Proposition~\ref{pro:perturbations}}

%\bibliographystyle{plain}

\bibliography{ifacconf}
\end{document}